\documentclass[lettersize,journal]{IEEEtran}

\usepackage[T1]{fontenc}

\usepackage{cite}

\ifCLASSINFOpdf
\else 
\fi
\usepackage{multirow}
\usepackage{romannum}
\usepackage[pdftex]{graphicx}
\graphicspath{ {./images/} } 
\usepackage{xcolor}
\usepackage{color} 
\usepackage{amsmath}
\usepackage[cmintegrals]{newtxmath}
\usepackage{subcaption}
\usepackage{indentfirst}
\usepackage{xcolor}
\usepackage{xpatch}
\usepackage{comment}

\begin{document}
\title{}
\title{MASSFormer: Mobility-Aware Spectrum Sensing using Transformer-Driven Tiered Structure}


\author{Dimpal~Janu,~\IEEEmembership{Member,~IEEE,} ~Sandeep~Mandia,~\IEEEmembership{ Member,~IEEE,} Kuldeep~Singh,~\IEEEmembership{Senior Member,~IEEE,} and
        Sandeep~Kumar,~\IEEEmembership{Senior Member,~IEEE} 
        ~
\thanks{D. Janu, and K. Singh  are with the Department
of Electronics and Communication Engineering, Malaviya National Institute of Technology Jaipur,
302017,India (e-mail: 2019rec9146@mnit.ac.in;kuldeep.ece@mnit.ac.in)}
\thanks{S. Mandia is with Thapar Institute of Engineering $\&$ Technology, Patiala, 147004, India (e-mail: sandeep.mandia@thapar.edu)}
\thanks{S. Kumar is with Central Research Lab, Bharat Electronics Ltd., Ghaziabad, 201010, India (e-mail: sann.kaushik@gmail.com)}
}        
\maketitle
\begin{abstract}
    In this paper, we develop a novel mobility-aware transformer-driven tiered structure (MASSFormer) based cooperative spectrum sensing method that effectively models the spatio-temporal dynamics of user movements.  Unlike existing methods, our method considers a dynamic scenario involving mobile primary users (PUs) and secondary users (SUs)and addresses the complexities
introduced by user mobility. The transformer architecture utilizes an attention mechanism, enabling the proposed method to adeptly model the temporal dynamics of user mobility by effectively capturing long-range dependencies within the input data. The proposed method first computes tokens from the sequence of covariance matrices (CMs) for each SU and processes them in parallel using the SU-transformer network to learn the spatio-temporal features at SU-level. Subsequently, the collaborative transformer network learns the group-level PU state from all SU-level feature representations. The  attention-based sequence pooling method followed by the transformer encoder adjusts the contributions
of all tokens. The main goal of predicting the PU states at each SU-level and group-level is to improve detection performance even more. We conducted a sufficient amount of simulations and compared the detection performance of different SS methods. The proposed method is tested under imperfect reporting channel scenarios to show robustness. The efficacy of our method is validated with the simulation results demonstrating its higher performance compared with existing methods in terms of detection probability, sensing error, and classification accuracy.  
\end{abstract}


\IEEEpeerreviewmaketitle

\section{Introduction}
\IEEEPARstart{T}{he} fifth generation (5G) mobile communication systems confront new requirements and challenges in terms of ultra-large range, massive connection, and ultra-low latency due to the exponential growth of mobile data traffic and the excessive connecting devices. 
The increasing use of wireless networks worldwide has caused a shortage of radio spectrum. Cognitive Radio (CR) networks offer a potential solution by allowing secondary users (SUs) to opportunistically access the spectrum allocated to primary users (PUs)  when it is not in use. Therefore, spectrum sensing (SS) has gained intense attention from academia in recent decades.  Numerous SS approaches have been proposed, with energy detection (ED) being the most commonly used method due to its low structure and minimal processing complexity \cite{kumar2018performance}.


\par Various deep/machine learning (DL/ML) methods have been used recently to enhance cooperative spectrum sensing (CSS) detection performance.
In \cite{thilina2013machine}, 
several ML-based SS methods have been proposed. A detailed analysis of ML-based approaches applied to CSS has been provided in \cite{janu2022machine}. This paper examines the strengths and limitations associated with these ML-based approaches, focusing on aspects such as performance improvement, variety of features used, applicable scenarios and complexity. There are certain limitations associated with ML-based approaches, including the necessity for manually crafted features during the ML model training process, as these features may not effectively capture the complexities of the real environment. Nowadays, DL-based methods have been applied to wireless communication demonstrating remarkable performance. Specifically, DL-based spectrum sensing has started to make a notable impact. 
Convolutional neural networks (CNNs) possess a robust ability to extract spatial features from data in matrix format. The covariance matrices (CMs) are regarded as a versatile statistical measure, encompassing correlation features of sensing signals of SUs. Consequently, CNNs have been widely applied in  SS  to acquire correlation features from CMs. 
 Liu \textit{et. al.} have designed CNN-based approaches in \cite{8761360} and \cite{liu2019deep}, where CMs were fed as input to derive the test statistics, and these two works achieved performance improvement in CSS. The SS methodology discussed in \cite{chen2020deep}  integrates the CNNs and short Time Fourier transform (STFT) to predict PU states leveraging time-frequency domain information. 
Graph convolutional neural networks (GCN) possess a robust capability to capture relationships among graph nodes by encoding the structural information of non-grid data. Additionally, it can concurrently incorporate graphs of variable sizes. Leveraging the distinctive ability of GCN, an SS method has been proposed in \cite{ janu2022graph} to solve the hidden node problem. In \cite{mehrabian2022cnn}, the authors used a CNN to provide a suitable alternative solution to the likelihood ratio test (LRT) under various general noises, which include Middleton class A, isometric complex symmetric $\alpha$ -stable, and isometric complex generalized Gaussian noise.  
\par The methodologies mentioned above leverage sensing data produced during the current sensing duration for the prediction of PU states while refraining from incorporating sensing data from preceding durations. 
 These methods utilize CNN architecture to learn the graphical features from CMs of sensed signals. Concurrently, researchers have also concentrated on capturing temporal information by leveraging historical data. 
An approach based on activity pattern aware spectrum sensing (APASS) has been developed in \cite{xie2019activity} in order to enhance the sensing performance even further. 
This method involves the simultaneous deployment of two parallel CNN structures, utilizing CMs generated from both current and past sensing data to learn both graphical and temporal features.  
A limited number of existing works in literature have applied the Long Short-Term Memory (LSTM) network, demonstrating its ability to learn temporal features through the utilization of historical sensing data across numerous sensing durations. In \cite{soni2020long}, an SS model that applies the LSTM network to capture temporal correlation features from sequential data of current and past sensing events has been proposed. Additionally, the model employed statistical insights of PU activity such as PU ON and OFF periods' durations as well as duty cycle to improve the sensing performance. 
Further, an LSTM-based SS approach has been developed in  \cite{chen2022cm} to learn temporal information from the CMs. Authors of \cite{yu2018spectrum} have developed a spectrum prediction method that utilizes the Taguchi method to optimize the hyperparameter of the LSTM network. 
Considering the single antenna SU scenarios, authors of  \cite{gao2019deep} and \cite{yang2019blind}    have developed an SS method employing a hybrid structure of CNN-LSTM to extract the spatio-temporal information from observations of single sensing intervals. A combination of 1-D CNN and LSTM network has been employed to learn the time and frequency domain features,  to find the presence of PU specifically in low signal-to-noise ratio (SNR) environments \cite{ke2019blind}.
In order to enhance the SS performance further, xie \textit{et. al.} have introduced a CNN-LSTM method in \cite{xie2020deep}. Initially, the detector employs CNN to capture spatial information from CMs derived from sensing signals. Subsequently, the features extracted from different sensing durations are inputted into the LSTM network to determine the PU activity pattern. An SS model utilizing CNN and LSTM network has been proposed in \cite{solanki2021deep}, where a 1-D signal vector is inputted to CNN to learn graphical features, and then learned features are provided as input to the LSTM network to capture the temporal features.     
\par The methodologies discussed above have utilized an LSTM network and a hybrid CNN-LSTM network to learn temporal information for predicting the PU states.
 While some LSTM-based CSS methods learn temporal features from observations collected during one sensing duration. However, such features are comparatively less precise than those obtained by extracting temporal features across several sensing periods to find the activity pattern of PU. The two activities, predicting PU states at individual SUs and at the fusion centre, occur simultaneously over time. 
However, the methods discussed above treat all sensing outcomes from participating SUs with equal importance,  which can misguide the PU activity prediction results by overstating the importance of irrelevant features. In general, these methods face challenges to effectively model long-range dependence in the input sequence and capture comprehensive spatio-temporal features among cooperating SUs. Moreover, all the above SS methods have assumed static SUs, and none of these methods have considered mobility scenarios. However, in real-world scenarios, PUs and SUs are mobile, and their mobility has a significant impact on detection performance. In dynamic wireless environments, accurately detecting PU activity patterns is challenging due to severe path loss and channel fading resulting from user mobility. 

\par Based on the above observation, we consider a mobile scenario having multiple SUs equipped with multiple antenna and propose a MASSFormer based method to effectively model the spatio-temporal dynamics of users by utilizing historical sensing data and exploiting user mobility patterns. Since transformer \cite{vaswani2017attention}, an attention mechanism based architecture can address the challenges that LSTM and RNN networks struggle with, particularly in effectively modeling long-range dependencies in input sequences. Vision transformer (ViT) \cite{dosovitskiy2020image}, a transformer based model, was first proposed for image classification and sparked considerable interest within the research community, leading to subsequent works and its extension to video vision transformer (ViViT) in \cite{arnab2021vivit}. Our proposed MASSFormer method is inspired by ViViT, which utilizes an SU-transformer network to learn relevant spatio-temporal features from the sensing outcomes of each SU over several sensing durations. Subsequently, a collaborative transformer network is utilized to model the SU-level spatio-temporal features of all participating SUs to learn group-level features to predict PU states. Attention across multiple levels of the transformer layers can effectively capture more relevant features from the individual SUs in a progressive manner. With the considered scenario, the CNN-LSTM method  \cite{xie2020deep} captures the temporal information from all SUs concurrently using a single LSTM, thereby neglecting temporal information at individual SU-level. Similarly, the $3$-D CNN \cite{tran2015learning} method fails to capture the spatio-temporal information at individual SU-level. 
\par We summarize the main contributions of the paper as follows:
\begin{itemize} 
    \item We develop a novel MASSFormer method that effectively models the spatio-temporal dynamics of user movements. The developed method first uses an SU-transformer network which uses the transformer encoder with attention mechanism to learn the spatio-temporal features of each SU to predict the PU states at SU-level and a collaborative transformer network to model the learned representations from SUs to predict the PU states at group-level.
    \item We adopt a practical system model having multiple SUs with multiple antennas. We address the challenges posed by the mobile users in real-world scenarios by accounting for varying levels of path loss and fading severity at each SU. We consider the impact of imperfect reporting channels between SUs and fusion centre to show the robustness of our approach.

    
    \item With extensive simulations, we compared and analyzed the detection performance of the proposed method with state-of-the-art methods. We have validated that the proposed method outperforms the state-of-the-art methods in considered mobile scenarios in terms of detection probability, sensing error, and classification accuracy.   
    
\end{itemize}

\section{System model}
We consider that a single PU with a single antenna and $S$ SUs with multiple antenna are distributed randomly in a predefined area of the CR network. In the considered mobile scenarios, it is assumed that users are moving with random velocity, resulting in dynamic changes to their locations over time, and their movements are independent of each other. At the beginning of each frame, each SU conducts SS and gathers $N$ observations at each antenna during $u$-th sensing intervals. We formulate two hypotheses regarding the states of PU, where $H_{1}$ and $H_{0}$ denote active state and inactive states, respectively. The signal $ y_{s}^{m}(n)$ from $m$-th antenna $s$-th SU is received at fusion centre represented as


\begin{equation}\label{HypothesesAtFusioncenter} 
    y_{s}^{m}(n) = \\
    \begin{cases}
   {h_{s,r}^{m}(n) (h_{s}^{m}(n)w(n) + \eta_{s}^{m}(n))}+ \eta_{c}(n) , & \mathcal{H}_{1} \\ 
     h_{s,r}^{{m}}(n)\eta_{s}^{{m}}(n)+\eta_{c}(n),    & \mathcal{H}_{0} 
    \end{cases}  \tag{1}
\end{equation} 
Where, PU signal is denoted as $w(n)$, $m=1,2,..., M$,  $s=1,2,...,S$ and $n=1,2,3,...,N$.  $\eta_{s}^{m}(n)$ and $h_{s}^{m}(n)$ denotes the noise signal and channel gain between PU and $m$-th antenna of $s$-th SU, respectively. The noise signal and channel gain received at fusion centre from 
$m$-th antenna of $s$-th SU is represented as $\eta_{c}(n)$ and $h_{s,r}^{m}(n)$, respectively.
It is assumed that signals traveling through sensing as well as reporting channel may undergo different amounts of fading, which is quantified by varying fading parameter values.
Channel gains of sensing channel and reporting channel are assumed to follow fading distribution over multiple sensing periods and during a particular sensing period it remains constant as sensing period is shorter than channel coherence time. 
 
\subsection{Mobility scenario}
\begin{figure}
    \centering
    \includegraphics[width=8cm,height=6cm]{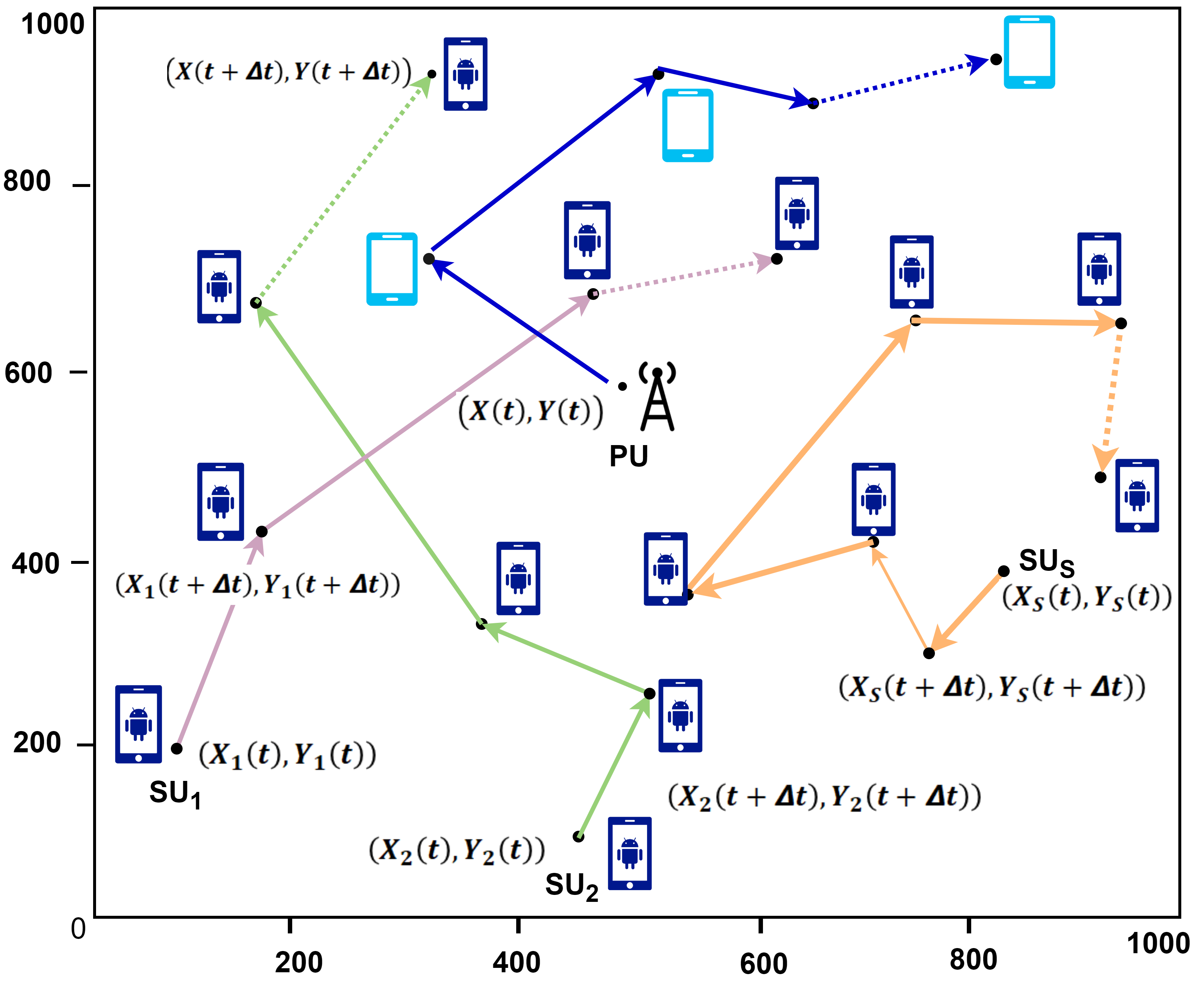}
    \caption{Movement pattern of PUs and SUs using RW Mobility model} 
    \label{fig:MobilityScenarios}
\end{figure}

Mobility scenarios involve the movement of both PUs and SUs within the network environment.
PUs mobility affects the prediction of PU states due to changes in PU activity patterns with their movements. The PU signals reaching SUs encounter varying levels of path loss and different channel conditions due to SUs' mobility. 
In this work, we consider the Random Waypoint (RW) Mobility Model \cite{camp2002survey} to find the movement patterns, trajectories, speed, direction, and location information of PUs and SUs. 
According to the RW mobility model, PU and SUs are initially distributed randomly in the simulation area. The users wait for pause time by staying in one location, and once this time expires, they choose a random destination in the simulation area and a speed that is uniformly distributed $[v_{min}, v_{max}]$.  The users move to the new location at the selected speed and wait for a pause time after arrival. Fig. \ref{fig:MobilityScenarios} shows the movement pattern of PU and SUs within a defined simulation area. Initially the position of moving user i.e. PU or SUs is represented as $(X,Y)$, after a time-interval $\Delta t$, the position is updated using the following formulas:
\begin{equation}
    X(t+\Delta t) = X(t)+v {\Delta t}\ast cos\theta(t)
    \tag{2} \label{xcord}
\end{equation}
\begin{equation}
    Y(t+\Delta t) = Y(t)+v {\Delta t} \ast sin\theta(t) \tag{3} \label{ycord}
\end{equation}
where speed $v$ is randomly selected between $[v_{min}, v_{max}]$, $\theta(t)$ denotes the direction at time $t$. The position traces of PU and SUs are calculated using equation (\ref{xcord}) and (\ref{ycord}).
\par The instantaneous signal-to-noise ratio experienced by  $s$-th SU is denoted as $\gamma_{s}= \frac{{|h_{s}|}^2Pt}{N_{0}B_{W}} =\frac{Pr_{s}}{N_{0}B_{W}}$. We assume that the transmit power of PU is fixed to $P_{t}$ and PU signals are transmitted through the channel whose bandwidth is $B_{W}$. 
The received PU power at SU $s$  at a distance $d_{s}(u)$ from the PU at time $u$ can be expressed in dB as 
\begin{equation*} 
    Pr_{s}(dB)=Pt(dB)- \{10log_{10}(\beta)+10\alpha log_{10}(d_{s}(u))\} \\ 
    \tag{4} \label{equation2}
\end{equation*} 
where $\alpha$ and $\beta$ represent path-loss exponent and path-loss constant respectively, and $N_{0}$ denotes noise power spectral density. 

\begin{figure*}[!h] 
    \centering
    \includegraphics[width=18cm,height=8cm]{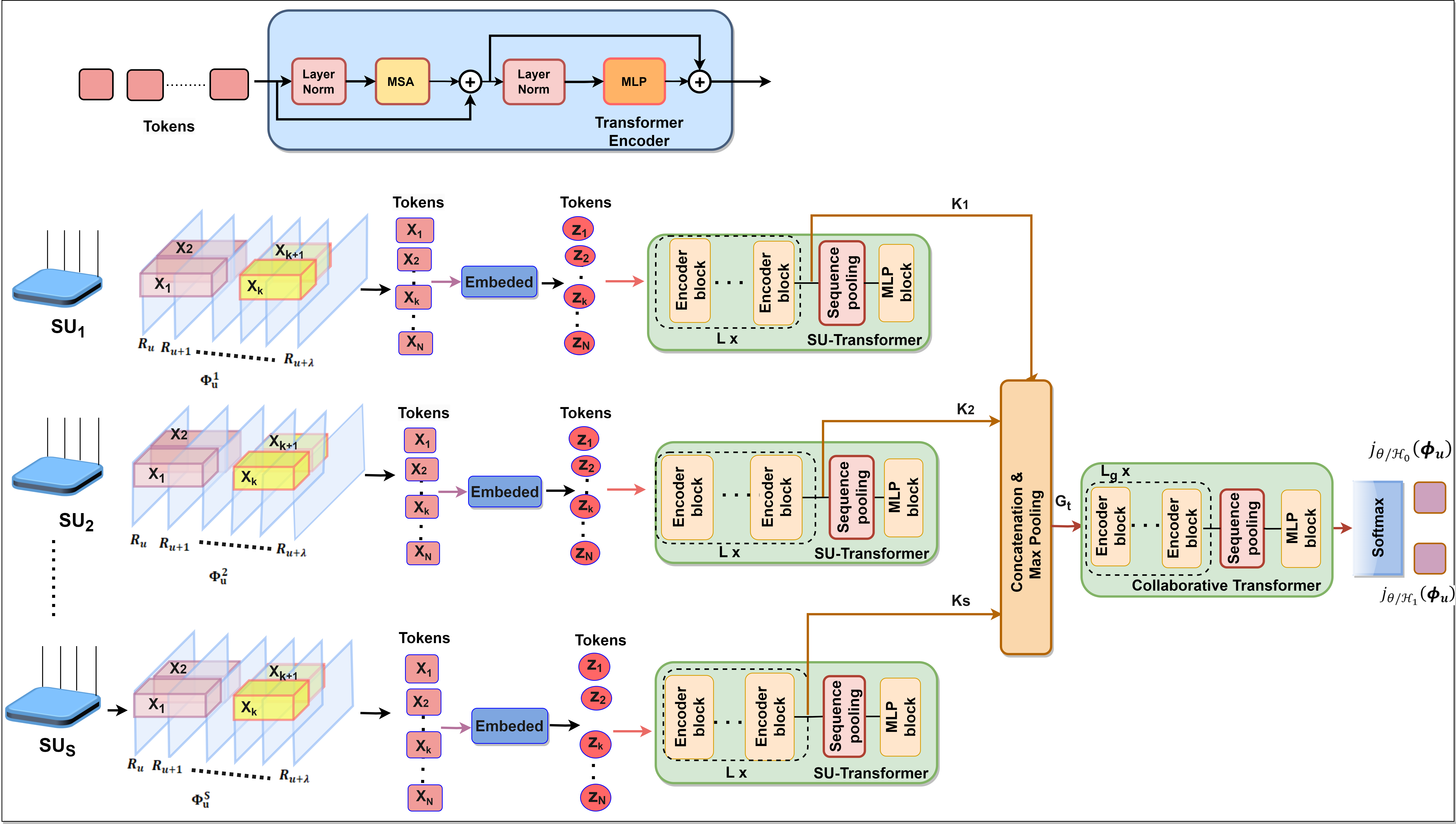}
    \caption{Architecture of MASSFormer} 
    \label{fig:TransformerbasedCSS}
\end{figure*} 
\subsection{Data Preprocessing}
The proposed model requires labeled training dataset collected in $U$ sensing durations. 
The raw sensing signal $ \boldsymbol{Y}_{u}^{s}$  collected during a single sensing duration is too large with dimension $M\times N$, to be directly considered as input for the neural network. Most of the existing SS methods utilized CMs as test statistics. Therefore, it is required to compress the sensing signal into CMs to construct the training set. 
The signal matrix $\boldsymbol{Y}_{u}^{s}$ composed of sensing signals from  $M$ antennas of SU $s$ to fusion centre  can be expressed as 
\begin{equation} \label{sensingsignal}
\boldsymbol{Y}_{u}^{s} = 
\begin{bmatrix}
y_{s}^1(1) & y_{s}^1(2)  &\cdots & y_{s}^1(N) \\
y_{s}^2(1) & y_{s}^2(2)  &\cdots & y_{s}^2(N) \\
\vdots  & \vdots  & \ddots & \vdots   &  \\
y_{s}^M(1) & y_{s}^M(2)  &\cdots & y_{s}^M(N)  \\ 
\end{bmatrix} \tag{5}
\end{equation} 
Let us construct the CM of the signal matrix as defined below 
\begin{equation} \label{CM}
 \boldsymbol{R}_{u}^{s} = \frac{1}{N} \boldsymbol{Y}_{u}^{s}{\boldsymbol{Y}_{u}^{s}}^H  \tag{6}
\end{equation}
\newline Now, construct the training set by arranging the CMs as defined as $\{ (\boldsymbol{R}^{s}_{1},b_{1}),  (\boldsymbol{R}_{2}^{s},b_{2}),(\boldsymbol{R}_{3}^{s},b_{3})...........(\boldsymbol{R}_{U}^{s},b_{U}) \}, $ where $U$ denotes total examples in the set. The class of $u$-th example is denoted as $b_{u} \in \{0,1 \}$, where the hypotheses $\mathcal{H}_{1}$ and $\mathcal{H}_{0}$ are selected when $b_{u}=1$ and  $b_{u}=0$, respectively.
To predict the PU states at SU-level, we concurrently analyze the CMs of sensed signals gathered at $s-$th SU during numerous sensing intervals. 
The dataset for  $s$-th SU is written as
\begin{equation} \label{SU_dataset}
   {\Phi}^{s}=
    \{ (\boldsymbol{\phi}_{1}^{s},b_{\lambda}),(\boldsymbol{\phi}_{2}^{s},b_{2\lambda}), ...,(\boldsymbol{\phi}_{u}^{s},b_{u\lambda}),.. 
    ,(\boldsymbol{\phi}_{U/\lambda }^{s},b_{U}) \} \tag{7}
\end{equation} 
where, $ \boldsymbol{\phi}^{s}_{u} = [ \boldsymbol{R}^{s}_{\lambda(u-1)+1},\boldsymbol{R}^{s}_{\lambda(u-1)+2},....,\boldsymbol{R}^{s}_{\lambda(u-1)+\lambda} ] $ and $\lambda$ denotes the length of input's temporal sequence. 
The $u$-th example in the set of $s$-th SU is denoted as $\boldsymbol{\phi}^{s}_{u}$. The complete dataset for  $s$-th SU is represented as $ {\Phi}^{s}$.
To generate data for all SUs, we follow the same method as described in equation (\ref{sensingsignal}), (\ref{CM}), and (\ref{SU_dataset}). Since every participating SU sends its sensing data to the fusion centre, the complete dataset constructed by collecting samples from all SUs is defined as


\begin{align*} \label{Complete_dataset} 
   \boldsymbol{\Phi} = 
    \{ (\boldsymbol{\phi}^{1}_{1},\boldsymbol{\phi}^{2}_{1},..,\boldsymbol{\phi}^{S}_{1},b_{\lambda}), (\boldsymbol{\phi}^{1}_{2},\boldsymbol{\phi}^{2}_{2}, ..,\boldsymbol{\phi}^{S}_{2},b_{2\lambda}) \\  ..,(\boldsymbol{\phi}^{1}_{u},\boldsymbol{\phi}^{2}_{u},..,\boldsymbol{\phi}^{S}_{u},b_{u\lambda})  ..,(\boldsymbol{\phi}^{1}_{U/\lambda },\boldsymbol{\phi}^{2}_{U/\lambda },..,\boldsymbol{\phi}^{S}_{U/\lambda},b_{U}) \}  \tag{8}
\end{align*} 

\section{MASSFormer based framework}
The equation (\ref{Complete_dataset}) denotes the complete dataset, where each sample in the set consists of data samples from all participating SUs. The data point  $ \boldsymbol{\phi}_{1}^{s}$ in equation (\ref{SU_dataset}), consists of a sequence of CMs computed for each of SU is represented as $\mathbb{R}^{\lambda \times M \times M} $.  Since, the transformer takes input in tokenized format, which is discussed in section \ref{tokenization}.

\subsection{Tokenization} \label{tokenization}
In this section, we discuss the tokenization process of ViViT, as we adopt a similar approach for tokenization in our methodology. A video sample is denoted as $X \in {\mathbb{R}^{T \times H \times W \times C}}$, where $T$, $H$, $W$, and $C$ referring to the temporal length, height, width and depth of the input respectively. 
The ViT-based model processes the $2$-D images to extract $N_{t}$ non-overlapping patches. In the context of videos, a sequence of spatio-temporal "tubes"  $x_{1},x_{2}......x_{N_{t}}$, 
$ {x} \in {\mathbb{R}}^{{t}\times {h} \times {w}}$ are extracted from the input volume as described in ViViT \cite{arnab2021vivit}. 
A linear operator $E$ is applied to the tubes to linearly project them to $1$-D tokens $ {z} \in \mathbb{R}^{N_{t} \times d}$, where $N_{t}=n_{t}n_{h}n_{w}$, $n_{t}=\lfloor \frac{T}{t} \rfloor, n_{h}=\lfloor \frac{H}{h} \rfloor$, and $ n_{w}=\lfloor \frac{W}{w} \rfloor$. To perform linear projection,  $3$-D convolution is used with the kernel size $(t,h,w)$ and the strides $(t,h,w)$ in time, height, and width dimensions, respectively. In ViViT, a learnable token $z_{cls}$ is prepended to the token sequence. However, to incorporate the information from all the tokens, we apply attention based sequence pooling operation to compute a 1-D feature vector from 2-D encoder output. As the self-attention mechanism in the transformer encoder is order agnostic, positional embedding $P \in \mathbb{R}^{N_{t} \times d}$ is added to tokens to preserve the positional information. Therefore, we extract tokens from the sequence of CMs constructed for each of the SUs as depicted in Fig. \ref{fig:TransformerbasedCSS}. 
The sequence of tokens for $s$-th SU is denoted as 
\begin{equation}\label{equation6} 
    {z_{s}}=[Ex_{1},Ex_{2},Ex_{3}......Ex_{N_{t}}+P] \tag{9}
\end{equation}

\subsection{Transformer encoder}
The SU-transformer network and collaborative transformer network are made up of several transformer encoder layers. The encoder layer consists of multi-head self-attention (MSA) and multi-layer perceptron (MLP) in sequence with skip connections. The sequence of tokens $z$ corresponding to each SU is processed by the SU-transformer network which consists of $L_{1}$ encoder layers. Each encoder layer is applied sequentially which consists of the following operations, 
\begin{equation}
   c^{l}=MSA(LN(z^{l-1}))+z^{l-1},
    \tag{10}
\end{equation}
\begin{equation}
    z^{l}=MLP(LN(c^{l}))+c^{l}
    \tag{11}
\end{equation} 
where LN is layer normalization \cite{ba2016layer}, MSA is multi-head self attention \cite{vaswani2017attention} and MLP \cite{hendrycks2016gaussian} consists of two dense layer separated by GeLU non-linearity. 
The self-attention aims to capture interactions among all $N_{t}$ token embeddings. 
The self-attention is computed as
\begin{equation} \label{self_attention}
	Z_{att}= \text{Softmax}\left(\frac{Q\left(K\right)^{T}}{\sqrt{d_{k}}}\right) \tag{12}
\end{equation}
The output of the self-attention block is obtained by multiplying $Z_{att} $ with the value matrix $V$.
MSA allows the model to concurrently attend to information from various representation subspaces at different positions. 
The output of MSA block is computed as:
\begin{equation}
	\text{MultiHead}(Q,K,V)= Concat(head_{1},head_{2}... head_{h_{d}})W^{O}, \tag{13}
\end{equation} 
where
\begin{equation}
	\text{head}_{i} = \text{Softmax}\left(\frac{QW^{Q_{i}}\left(KW^{K_{i}}\right)^{T}}{\sqrt{d_{k}/h_{d}}}\right)VW^{V_{i}} \tag{14}
\end{equation} where $W^{Q_{i}}$ $\in$ $R^{d \times d_{q}}$,  $W^{K_{i}}$ $\in$ $R^{d \times d_{k}}$ and  $W^{V_{i}}$ $\in$ $R^{d \times d_{v}}$, and  $W^{O}$ $\in$ $R^{h_{d}d_{v}\times d}$, $d_{v}=d_{k}=d/h_{d}$.  $i$ denotes number of heads ranging from $1$ through $h_{d}$ and $h_{d}$ being the total number of heads in MSA block.
\subsection{Architecture of MASSFormer} 
In this section, we provide a brief introduction to the proposed MASSFormer method. The main objective of the proposed architecture is to learn the collective PU states by selectively extracting the most relevant features from the sensing data of all cooperating SUs.
The proposed MASSFormer model is divided into two components i.e. the SU-transformer network and collaborative transformer network to predict PU states by extracting spatio-temporal features at individual SU-level and at group-level respectively. 
The proposed MASSFormer architecture is depicted in Fig. \ref{fig:TransformerbasedCSS}, where SU-transformer network is used to predict the SU-level PU states by extracting spatio-temporal features from the sequence of tokens extracted from a series of CMs of a $s$-th SU. Further, collaborative transformer network is used to model the output representations of all SUs to predict the PU states at the group-level. Attention from various levels of a transformer can extract contributing features from the participating SUs in a progressive manner.  

To predict the PU states, tokens computed from samples of each SU are fed to SU-transformer network in different parallel pipelines. SU-transformer network consists of  $L_{1}$  encoder layers,  followed by sequence pooling as described in section \ref{seqpool} and MLP block to predict the PU states at SU-level. We concatenate the output feature representations $K_{s}$ of each $s$-th SU.
\begin{equation*} \label{Concatenatedoutput}
    G_{t}= K_{1} \diamond K_{2}...\diamond K_{s}...\diamond K_{S}  \tag{15}
\end{equation*} 

Further, max pooling operation is performed over the feature $K_{s}$ of all SUs to get output representation $G_{t}$ provided in equation (\ref{Concatenatedoutput}).
Instead of spatial pooling (like typical max pooling in CNNs, which pools over spatial dimensions), this operation pools over the dimension representing different SU and combines their predictions by selecting the maximum value for each position across SU dimension. The output features $G_{t}$ is fed to collaborative transformer network which consists of $L_{2}$ transformer encoder layer, followed by sequence pooling  and MLP block, to extract the group-level features from all the participating SU. Finally, the output features are fed to Softmax layer for  predicting the activity states of PU. 
 \subsection{Sequence pooling}\label{seqpool}
 For the prediction of class label, ViViT \cite{arnab2021vivit} forwards a learnable class token through the encoder layers and later to the MLP for classification. In contrast, we use sequence pooling, first proposed by Hassani \textit{et. al.} \cite{hassani2021escaping} to extract representation. Sequence pooling, an attention-based method, assigns weights to sequential embeddings, transforming the input sequence into a vector representation. This involves assigning importance weights to the processed data from transformer encoder. The motivation is rooted in the dispersion of information across all tokens in the output sequence, necessitating aggregation by assigning suitable weights to each token. Sequence pooling involves the transformation of sequence as $T:$ $\mathbb{R}^{N_{t}\times d} \mapsto \mathbb{R}^{d}$. Given the output $z \in \mathbb{R}^{N_{t} \times d}$ of encoder is passed to linear layer $g(z) \in \mathbb{R}^{N_{t}\times 1}$, and softmax activation to the token sequence as follows 
\begin{equation}\label{weight}
    Weigths= softmax(g(z)^T) \in \mathbb{R}^{1 \times N_{t}} \tag{16}
\end{equation} 
Hence, we calculate importance weights for each input token as described in equation (\ref{weight}), and further, these weights are assigned to each token as described in equation (\ref{outputofseqpool}). The computed weights are employed to adjust the contributions of the tokens through a weighting operation as follows
\begin{equation}\label{outputofseqpool}
    z_{seq}=Weigths \times z \in \mathbb{R}^{1\times d} \tag{17}
\end{equation} 
$z_{seq}$ is passed to MLP block of the SU-transformer network and collaborative transformer network to detect activity states of PU at SU-level and group-level respectively.
\subsection{Network training} 
The developed model is trained in two distinct stages. In the initial stage, the SU-transformer network is trained end-to-end using a training dataset comprising sequences of CMs converted into tokens and associated labels. This process aims to extract spatio-temporal features at SU-level. After the SU-transformer network gets trained, the output features, $K_{s}$, extracted for each SU during the initial stage are concatenated and further, a max pooling operation is applied to obtain pooled feature representation denoted as $P_{t}$. In the subsequent stage, the pooled features from multiple sensing periods are inputted into the collaborative transformer network which comprises various encoder layers, followed by sequence pooling and an MLP block.

The resulting output features are fed to the Softmax layer for the prediction of probabilities associated with hypotheses $\mathcal{H}_{0}$ and $\mathcal{H}_{1}$. The $u$-th example in the set contains data samples from each SU  is denoted by symbol ${\Phi}_{u}$, where ${\Phi}_{u}=(\phi_{u}^{1},\phi_{u}^{2}...\phi_{u}^{S})$. Normalized output of the developed model is represented as $[  j_{{\theta}/\mathcal{H}_{0}}(\boldsymbol{\Phi}_{u}), j_{{\theta}/\mathcal{H}_{1}}(\boldsymbol{\Phi}_{u}) ]$, where $j_{{\theta}/\mathcal{H}_{0}}(\boldsymbol{\Phi}_{u})+j_{{\theta}/\mathcal{H}_{1}}(\boldsymbol{\Phi}_{u})=1$.
For a new data point $\boldsymbol{\Phi}_{u}$, the trained model predicts a probability associated to either Hypothesis $\mathcal{H}_{0}$ and $\mathcal{H}_{1}$. 
$j_{{\theta}/\mathcal{H}_{i}}(\boldsymbol{\Phi}_{u})$ denotes the class probability of hypothesis $\mathcal{H}_{i}$. %
Based on this, the main goal of training is to maximize the likelihood as
\begin{align*}  
\prod \limits _{u = 1}^{U/\lambda} {{{({j_{\theta |{\mathcal{H}_{1}}} }({\boldsymbol{\Phi}_{u}))}^{b_{u\lambda}}}{{({j_{\theta |{\mathcal{H}_{0}}} }(\boldsymbol{\Phi}_{u}))}^{1 - {b_{u\lambda}}}}}}.\tag{18}
\end{align*} 
This is equivalent to minimizing the cost function, a cross-entropy loss function, which is defined as logarithm of the likelihood function and normalizing it with the number of training points. 
\begin{align*}  
 L({\theta})= -\frac{1}{ U/\lambda} \sum \limits _{u = 1}^{U/\lambda} {{{b_{u\lambda}}{\log j_{\theta |{\mathcal{H}_{1}}} }({\boldsymbol{\Phi}_{u})}} +{(1 - {b_{u\lambda})}}{{{\log j_{\theta |{\mathcal{H}_{0}}} }(\boldsymbol{\Phi}_{u})}}} \tag{19}
\end{align*}    
To achieve the maximum likelihood, the cross-entropy loss function is minimized to obtain the optimal parameters as 
\begin{equation*} 
{\theta ^{*}} = \arg \min \limits _\theta L(\theta).\tag{20}
\end{equation*} 
While training the developed model, Back propagation (BP) algorithm is used to compute gradients of the loss function, and applying Adam optimizer to optimize the model parameters to get well-trained network. The output of trained model is denoted as  $j_{\theta^{*} |{\mathcal{H}_{0}}} (\boldsymbol{\Phi}_{u})$ and $j_{\theta^{*} |{\mathcal{H}_{1}}} (\boldsymbol{\Phi}_{u})$. According to Neyman-Pearson (N-P) theorem, the optimal test statistics is likelihood ratio is represented as 
\begin{equation*} 
\beta_{MASSFormer}(\boldsymbol{\Phi}_{u}) = \frac { j_{\theta |{\mathcal{H}_{1}}} (\boldsymbol{\Phi}_{u})}{ j_{\theta |{\mathcal{H}_{0}}} (\boldsymbol{\Phi}_{u})}.\tag{21}
\end{equation*} 
Upon the arrival of a new data point $\boldsymbol{\Phi}_{u}$, the model output is used to determine the activity states of PU. However, this approach does not provide control over the false alarm probability $p_{fa}$. Therefore, we calculate the detection threshold $\gamma$ using Monte-Carlo method that ensures the desired false alarm probability $p_{fa}$ is achieved. 
We collect the sequence of CMs corresponding to hypothesis $\mathcal{H}_{0}$ to build a set 
$ \boldsymbol{\Phi}_{\mathcal{H}_{0}}=  \{ \Tilde{\boldsymbol{\Phi}}_{1},...., \Tilde{\boldsymbol{\Phi}}_{\Tilde{Q}} \}$, where total examples in set $\boldsymbol{\Phi}_{\mathcal{H}_{0}}$ denoted by $\Tilde{Q}$. Now, we fed the noisy data samples to the MASSFormer and get the expression of test statistics under $\mathcal{H}_{0}$ i.e.  $\beta_{MASSFormer|{\mathcal{H}_{0}}}$.
Further, reorganize the all these $\beta_{MASSFormer|{\mathcal{H}_{0}}}$ to construct a set $\tilde {\boldsymbol{\Phi}}_{MASSFormer|\mathcal{H}_{0}}$ in ascending order such as 
$\forall 1 \le p \le q \le \tilde Q,$
\begin{equation*}  
\quad {\beta_{MASSFormer|{\mathcal{H}_{0}}}}\left ({{{\tilde {\boldsymbol {\Phi }} _{p}}} }\right) \le {\beta_{MASSFormer|{\mathcal{H}_{0}}}}\left ({{{\tilde {\boldsymbol {\Phi }} _{q}}} }\right) \tag{22}\end{equation*}
Lastly, we compute the detection threshold with constraint of $p_{fa}$  as 
\begin{equation*} \gamma = { \tilde {\boldsymbol{\Phi}}_{MASSFormer|\mathcal{H}_{0}} }\left({round\left ({{\tilde Q\left ({{1 - {p_{fa}}} }\right)} }\right)}\right).\tag{23}\end{equation*} 
The function $round(.)$ provides the nearest integer value of a given number. For a test data, a decision about the activity states of PU can be made as follows
 
\begin{equation*}  
\gamma \mathop \lessgtr \limits _{\mathcal{H}_{0}}^{\mathcal{H}_{1}} \beta_{MASSFormer}(\boldsymbol{\Phi}_{u}).\tag{24} 
\end{equation*}

\section{Numerical Results} 
\subsection{Simulation environment and Hyperparameters} 
We consider the simulation area of $1000$m $\times$ $1000$m area, and they move with a velocity randomly chosen between $[20,25]$ m/sec, resulting in dynamic changes in their positions over time. The pause time is chosen as $1$msec.
This setup allows us to assess the feasibility of the proposed model under conditions of user mobility. 
We assume that PU signals are independent and identically distributed (i.i.d.) Gaussian Random vector with zero mean and signal variance $\sigma_{w}^{2}$. We have chosen values of $N=100$, $S=3$, and $M=15$, length of temporal sequence $\lambda=20$ randomly. 
PU signals are assumed to be sent on a channel having bandwidth $B_{W}$ is $10$MHz. 
Furthermore, we assume that $P_{t}=200mW$,  $\beta=10^{3.453}$ and $\alpha=3.8$, and noise power density is $N_{0}=-150dBm/Hz $. We assumed that the noise signal  represented by $\eta^m_{s}(n)$ and $\eta_{c}(n)$  follow an i.i.d Gaussian random vector with zero mean and noise variance  $\sigma_{\eta}^2$. To address the challenges occurring due to user mobility in real-world scenarios, we assumed that PU signals reaching from multiple SUs to fusion centre may experience different path loss and fading severity. Consequently, we assumed that channel gains $h_{s}^{m}(n)$ and $h_{s,r}^{m}(n)$ follow  Rayleigh distributions and fading severity is modeled by adjusting the fading parameter. The received signal power is calculated using equation (\ref{equation2}), which incorporates path-loss between PU and each SU. The hyperparameters used for our proposed architecture are provided in Table \ref{tab:Parameters and Dataset settings}. The number of training and testing points used for training and evaluating the proposed models are $104,000$ and $15,000$. The number of epochs and batch size are considered to be $100$ and $16$.

\setlength{\arrayrulewidth}{0.3mm} 
\setlength{\tabcolsep}{3pt} 
\renewcommand{\arraystretch}{1.2}
\begin{table}[t] 
\centering
\caption{HYPER-PARAMETERS  SETTINGS} 
\label{tab:Parameters and Dataset settings}

\begin{tabular}{lll}
\hline
\multicolumn{3}{l}{\textbf{Parameter  settings}}                                                                                                                                                         \\ \hline
\multicolumn{3}{l}{Input shape:- Dimension (20,16,16,3)}                                                                                                                                          \\ \hline
\multicolumn{3}{l}{Patch/token size: Dimension (20,1,1)}                                                                                                                                              \\ \hline
Parameter description     & \begin{tabular}[c]{@{}l@{}}SU-Tranformer \\ Network\end{tabular}                     & \begin{tabular}[c]{@{}l@{}}Collaborative \\ Transformer Network\end{tabular} \\ \hline
3D convoultion            & \begin{tabular}[c]{@{}l@{}}Kernel size:24@(20,1,1)\\ stride: Patch size\end{tabular} & -                                                                          \\ \hline
\begin{tabular}[c]{@{}l@{}}Linear layer \\projection of input \end{tabular}  & & \\ \hline
Projection dimention      & 24                                                                                   & 24                                                                           \\ \hline
No. of Heads              & 4                                                                                    & 4                                                                            \\ \hline
No. of encoder layers & 5                                                                                    & 4                                                                            \\ \hline
Transformer units         & {[}48,24{]}                                                                          & {[}48,24{]}                                                                  \\ \hline
MLP head units            & {[}128,64{]}                                                                         & {[}128,64{]}                                                                 \\ \hline
Learning Rate             & $1e^{-5}     $                                                          & $1e^{-5}$                                                       \\ \hline
Total parameters          & $42667$                                                                                & $31747 $         \\ \hline  
\end{tabular}
\end{table}
\setlength{\arrayrulewidth}{0.3mm} 
\setlength{\tabcolsep}{5pt} 
\renewcommand{\arraystretch}{1.2}
\begin{table}[t]
\centering
\caption{FLOPs for MASSFormer} 
\label{tab: FLOPs for MASSFormer}
\begin{tabular}{lll}
\hline
Layers Description                                                       & \begin{tabular}[c]{@{}l@{}}FLOPs\\ (SU-Transformer)\end{tabular} & \begin{tabular}[c]{@{}l@{}}FLOPs (Collaborative\\ Transformer)\end{tabular} \\ \hline
Patch Embedding                                                          & 245,760                                                           & 576                                                                          \\ \hline
MSA                                                                      & 1,376,256                                                         & 1,376,256                                                                    \\ \hline
MLP in Encoder                                                           & 1,179,648                                                         & 1,179,648                                                                    \\ \hline
Layer Normalization                                                      & 24,576                                                            & 24,576                                                                       \\ \hline
\begin{tabular}[c]{@{}l@{}}Total FLOPs for \\ Encoder Layer\end{tabular} & \begin{tabular}[c]{@{}l@{}}5 $\times$ (1,376,256 +\\ 1,179,648 + 24,576)\end{tabular} & \begin{tabular}[c]{@{}l@{}}4 $\times$ (1,376,256 +\\ 1,179,648 + 24,576)\end{tabular} \\ \hline
Sequence Pooling                                                         & 12,864                                                            & 12,864                                                                       \\ \hline
MLP Head                                                                 & 11,392                                                            & 11,392                                                                       \\ \hline
Total FLOPs                                                              & 13,172,416                                                        & 10,346,752                                                                   \\ \hline
MASSFormer                                                               & \begin{tabular}[c]{@{}l@{}}Params: 74,414\\ FLOPs: 23,519,168\end{tabular} & \\ \hline
\end{tabular}
\end{table}

\setlength{\arrayrulewidth}{0.3mm} 
\setlength{\tabcolsep}{8pt} 
\renewcommand{\arraystretch}{1.5}
\begin{table}[t]
\centering
\caption{FLOPs for CNN-LSTM} 
\label{tab: FLOPs for CNN-LSTM}
\begin{tabular}{lll}
\hline
Layers Description  & Parameters value    & FLOPs     \\ \hline
Input:              & (20,16,16,3)        &           \\ \hline
Conv. Layer         & Kernel: 32 @(3,3) & 8,847,360 \\ \hline
Max-pooling layer   & Kernel (2,2) stride:(1,1)                  & -         \\ \hline
Global Avg. Pooling & -                   & -         \\ \hline
Dense layer 1       & dimension: 128      & 163,840   \\ \hline
LSTM                & Units:32            & 819,200   \\ \hline
Dense layer 2       & Unit: 2             & 128       \\ \hline
Total FLOPs and Params        &  Total Params: 25,794                 &  9,830,400 \\ \hline 
\end{tabular}
\end{table}
\setlength{\arrayrulewidth}{0.3mm} 
\setlength{\tabcolsep}{8pt} 
\renewcommand{\arraystretch}{1.5}
\begin{table}[t]
\centering
\caption{FLOPs for 3D CNN} 
\label{tab: FLOPs for 3CNN}
\begin{tabular}{lll}
\hline
Layers Description  & Parameters value    & FLOPs     \\ \hline
Input:              & (20,16,16,3)        &           \\ \hline
Conv. Layer         & Kernel: 32$@$(3,3,3) & 26,542,080 \\ \hline
Max-pooling layer   & Kernel (2,2,2) stride:(1,1,1)                  & -         \\ \hline 
Conv. Layer         & Kernel: 24 $@$(3,3,3) & 177,292,800 \\ \hline 
Max-pooling layer   & Kernel (2,2,2) stride:(1,1,1)                  & -         \\ \hline 
Global Avg. Pooling & -                   & -         \\ \hline
Dense layer 1       & dimension: 64      & 1536   \\ \hline

Dense layer 2       & Unit: 2             & 128       \\ \hline
Total FLOPs         &  Total Params: 25,114  & 203,836,544   \\ \hline 
\end{tabular}
\end{table}
\begin{figure}
    \centering
    \includegraphics[width=8cm,height=6cm]{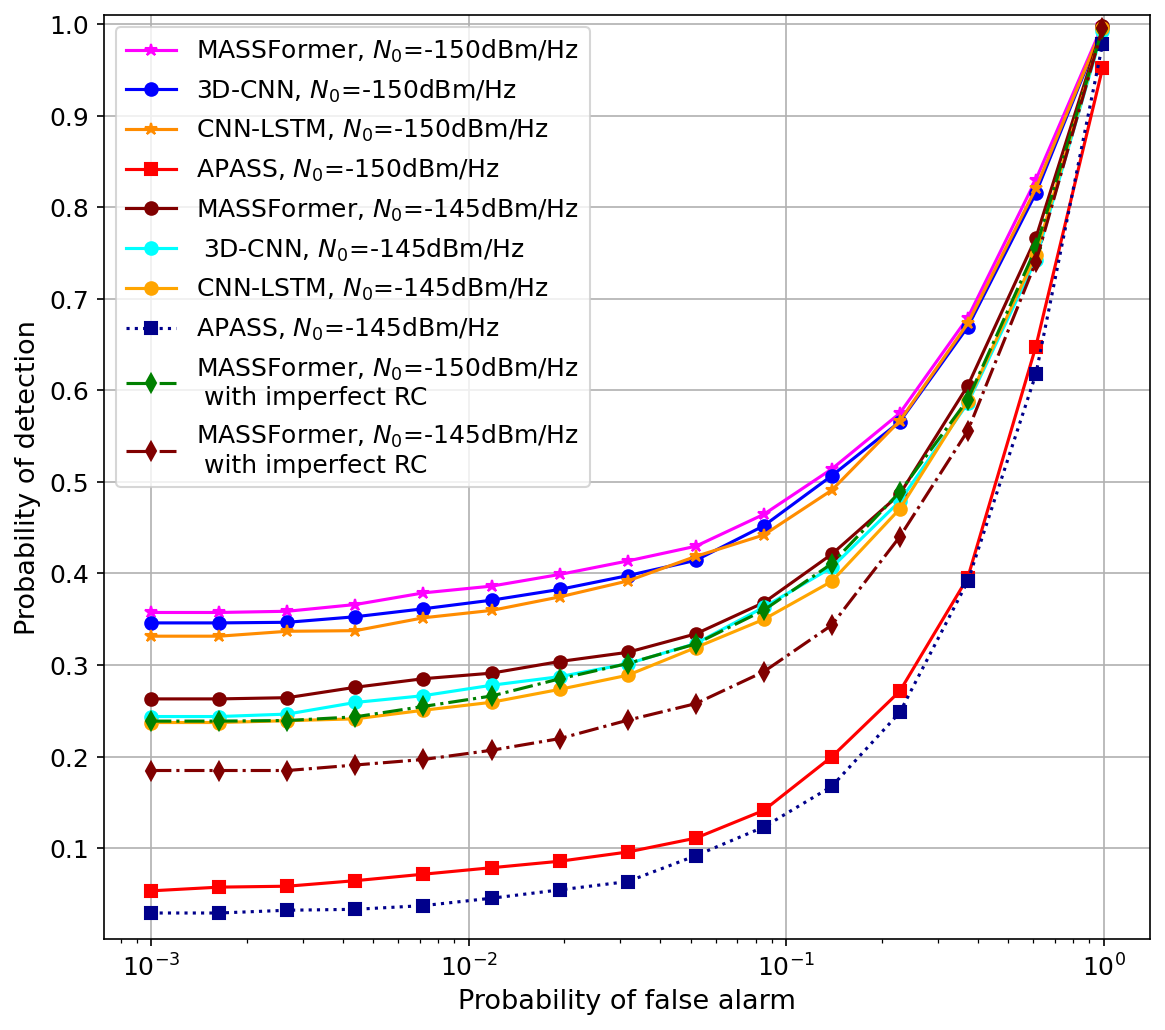}
    \caption{ROC curve for various SS methods} 
    \label{fig:ROC_curve}
\end{figure} 
\begin{figure}
    \centering
    \includegraphics[width=8cm,height=6cm]{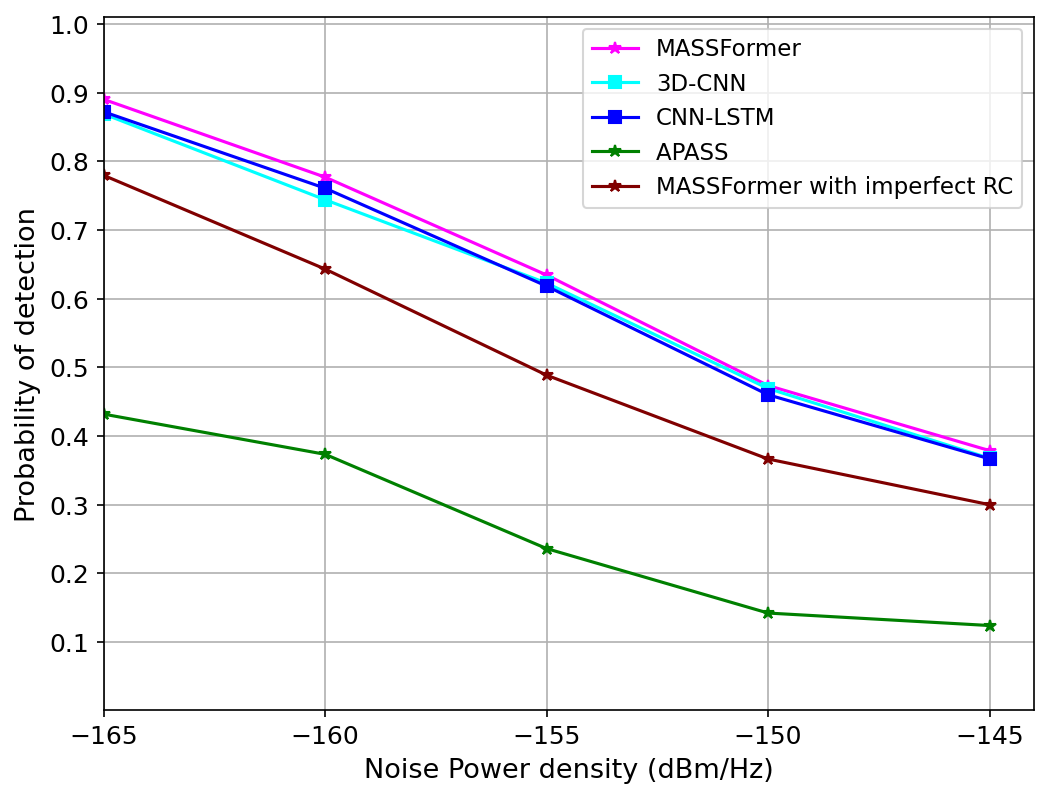}
    \caption{Probability of detection vs noise power density ($p_{fa}=0.09$)} 
    \label{fig:Pd_vs_SNR}
\end{figure} 

\subsection{Simulation Results} 
We conducted comprehensive simulations to illustrate the efficacy of the proposed MASSFormer method. We evaluated the detection performance in terms of receiver operating characteristics (ROC) curve, probability of detection ($P_{d}$) vs noise power density $(N_{0})$, and sensing error vs noise power density. We compared and analyzed the performance of MASSFormer method with existing methods such as CNN-LSTM \cite{xie2020deep}, APASS \cite{xie2019activity}, and $3$D CNN \cite{tran2015learning} methods. 
We rigorously evaluated the performance of the proposed method under various scenarios, including imperfect sensing as well as both perfect and imperfect reporting channels, to thoroughly test its robustness. 
We calculated the $P_{d}$ values at different detection thresholds computed at different values of $p_{fa}$ to evaluate the detection performance. 
Fig. \ref{fig:ROC_curve} shows the ROC curve of the proposed MASSFormer method along with existing methods at two different noise power density values such as $N_{0}=-150dBm/Hz$ and $N_{0}=-145dBm/Hz$. 

The results show that the proposed MASSFormer method performs better as compared to methods CNN-LSTM, APASS, and $3$D CNN in terms of detection performance.
\begin{figure} 
    \centering
    \includegraphics[width=8cm,height=6cm]{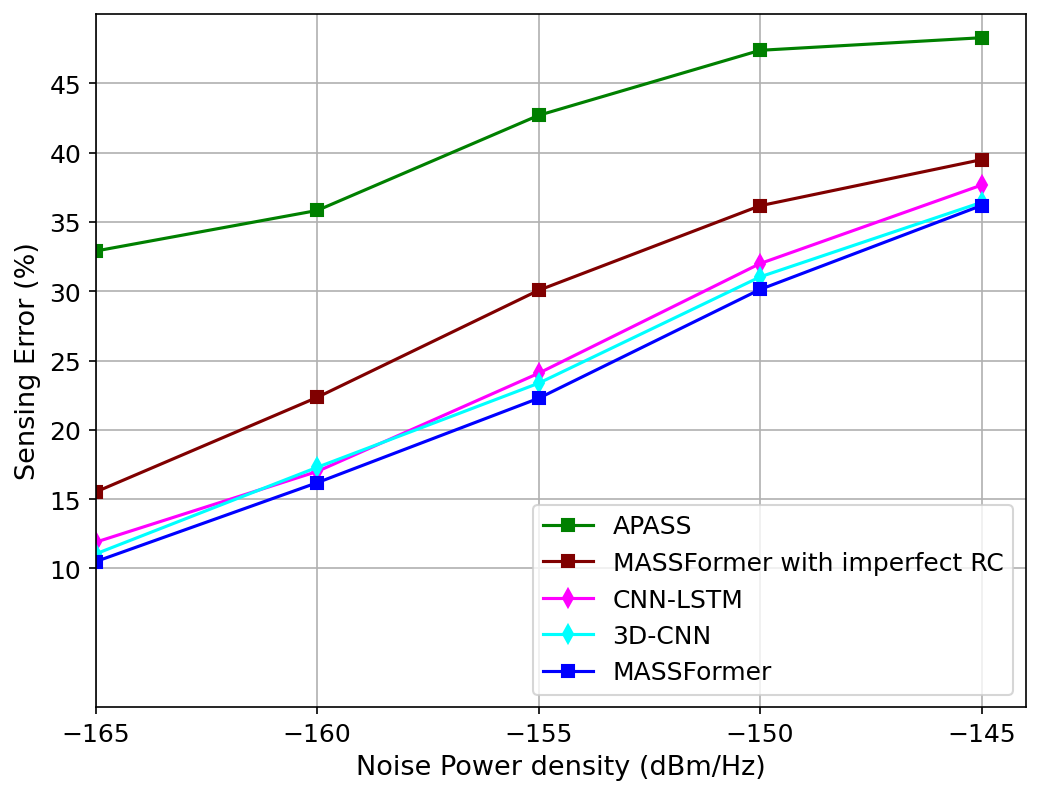}
    \caption{Sensing error vs noise power density} 
    \label{fig:SE_vs_SNR}
\end{figure} 
\begin{figure}
    \centering
    \includegraphics[width=8cm,height=6cm]{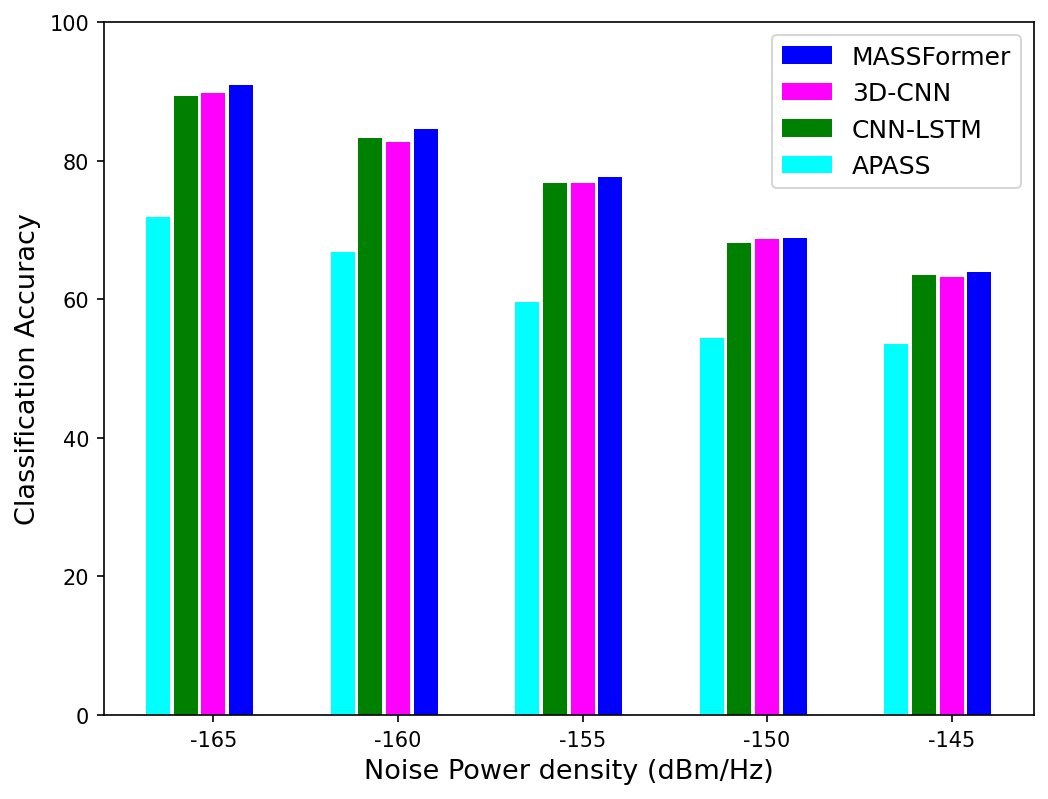}
    \caption{Classification Accuracy vs noise power density} 
    \label{fig:CA_vs_SNR}
\end{figure}
In accordance with IEEE 802.22 standard, the maximum acceptable value of false alarm probability is $(P_{f}\leq 0.1)$. 
Therefore, we provide a plot of $P_{d}$ versus noise power density at a fixed value of false alarm probability $P_{f}=0.09$ in Fig. \ref{fig:Pd_vs_SNR}. 
From Fig. \ref{fig:ROC_curve} and Fig. \ref{fig:Pd_vs_SNR}, we observe that the detection probability decreases as the noise power density increases.
The sensing error, which is defined as the mean of the probability of miss detection and probability of false alarm, another performance metric, is used to evaluate the performance of the proposed method. 
 

Fig. \ref{fig:SE_vs_SNR} depicts a plot of sensing error vs noise power density for different methods. 
We observe that the sensing error increases with increasing noise power density, reflecting inaccuracies in individual SUs' sensing outcomes. After analysis, we find that the proposed method achieves lower sensing error than the existing methods, highlighting its benefits.
Fig. \ref{fig:CA_vs_SNR} depicts the plot of classification accuracy vs noise power density for different methods. After a comprehensive analysis of the results, it is evident that the proposed MASSFormer method outperforms the existing methods, exhibiting superior detection performance in terms of detection probability and classification accuracy. 


\section{Computational Complexity Analysis}
Although the proposed method achieves improved detection performance, the computational complexity of detection methods should be analyzed for a fair comparison. Since training of neural networks can be performed offline, the paper focused on the computational complexity of inference after the model is deployed. The computational complexity to  convert the sensing signal $\boldsymbol{Y}_{u}^{s} $ into the CM $\boldsymbol{R}_{u}^{s}$ is $\text{O}(M^{2}N)$. Let $l$ and $L_{CNN}$ represent present convolutional layer's index and total number of convolutional layers.
We define the computational complexity of the convolutional layer as $ {O}\left ({\sum _{l=1}^{L_{CNN}} n_{f,l-1} \cdot n_{s,l}^{2} \cdot n_{f,l} \cdot  m^{2}_{s,l} }\right)$,
where $ n_{f,l}$ and $ n_{f,l-1}$ show the convolutional kernels count of $l$th and $(l-1)$th layer.
We define the spatial size of kernel current layer and the resulting output feature map as $n_{s,l}$ and $m_{s,l}$ respectively. With the CNN stride to be $1$, we have $m_{s,1}=M$ and the total number of SUs becomes the number of input channel i.e. $n_{f,0}=S $. The complexity of single convolutional layer is increased by the temporal length $\lambda$ since every sample has $\lambda$ CMs in the dataset.
The  complexity of LSTM network is denoted as ${O} \left(4\lambda n_{l} \left( n_{f,1}+n_{l}\right) \right)$, where the dimension of the each gate of LSTM cell are represented by $n_{l}$.
Therefore, complexty of CNN-LSTM method, which consists of Convolutional layers, Max-pooling layers, Global average pooling layer, and two dense layers, is provided below 
\begin{multline*} \label{compelxityy} 
    {O}\left(\lambda M^{2}n_{f,1} {n_{s,1}^{2}S} +
    {4\lambda n_{l1} (n_{f,1}+n_{l}})+{n_{f,1}}{n_{fc1}+n_{l}n_{fc2}}\right)     \tag{25} 
\end{multline*}
where, $n_{fc1}$ and $n_{fc2}$ denotes the dimensions of the first and second dense layer respectively, 
The computational complexity of $3$-D CNN is computed as below 

\begin{multline*}
    {O}\left(\lambda M^{2}( n_{f,1}{m_{s,1}^{3}} S + n_{f,1} n_{f,2} {m_{s,2}^{3}})+ n_{f,2} d_{fc1} +d_{fc1} d_{fc2}  \right)  \tag{26} 
\end{multline*}
where $n_{f,1} $, and $n_{f,2}$ denotes the number of filters of first and second convolutional layer respectively. We define the spatial size of the filter in the first layer and second as $m_{s,1}$ and $m_{s,2}$, respectively.
 $d_{fc1}$ and $d_{fc2}$ denote the dimensions of first and second dense layer respectively.  
\par The computational complexity of the proposed MASSFormer method is 
\begin{multline*}
     {O}\left( L_{1}h_{att1}M^{2}\lambda d_{emb1}+L_{1}Md_{fc1}\right) +\\
     {O}\left( L_{2}h_{att2}M^{2}\lambda d_{emb2}+L_{2}Md_{fc2} \right)  \tag{27} 
\end{multline*}
where $L_{1}$ and $L_{2}$  denote the number of transformer layers,  $h_{att1}$ and $h_{att2}$  number of attention heads in SU-transformer and collaborative transformer network respectively. $d_{emb1}$ and $d_{emb2}$ denote the projection dimensions, $d_{fc1}$ and $d_{fc2}$ denote the dimension of the dense layer of MLP block in SU-transformer and collaborative transformer network respectively.

\par The computational complexity of these methods is also computed in terms of multiply accumulates (MACs) or floating point operations (FLOPs) as provided in Table \ref{tab: FLOPs for MASSFormer}, \ref{tab: FLOPs for CNN-LSTM}, and \ref{tab: FLOPs for 3CNN}. From the analysis, it is concluded that the FLOPs required for the proposed MASSFormer method are less as compared to the $3$D CNN method and higher than the CNN-LSTM method. 
We measured the time needed for data preparation and model inference. We calculated the time for preprocessing the data for $1000$ samples, where each sample comprises of $\lambda=20$ CMs. After averaging, the time required for one data sample is $0.099$ msec.
For model inference, the CNN-LSTM, 3-D CNN, and APASS detector had inference times of $0.17$, $0.65$, and $0.87$ msec, respectively. 
Our proposed MASSFormer method requires $2.46$ msec for inference.
According to the IEEE $802.22$ standards, SUs are required to evacuate the spectrum within $2$ seconds when a PU becomes active. Therefore, our model detects the PU state within $2.46$ msec satisfying the real-time latency constraint. 


\section{Conclusion}
In this work, we proposed a MASSFormer method to predict the PU states in mobile scenarios. Since the prediction of PU state at SU-level and group-level, both events are occurring simultaneously over time. Therefore, inspired by this, a model is developed that uses SU-transformer network to predict PU states at SU-level and collaborative transformer network to predict PU states at group-level by modeling the spatio-temoral dynamics of movements of all contributing SUs. From simulations results, it is evident that our MASSFormer method outperforms existing methods in terms of $P_{d}$ demonstrating superior sensing performance.



\ifCLASSOPTIONcaptionsoff
  \newpage
\fi

\bibliographystyle{IEEEtran}
\bibliography{bibliography.bib}

\begin{IEEEbiography}[{\includegraphics[width=1in,height=1.25in,clip,keepaspectratio]{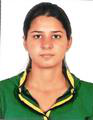}}]{Dimpal Janu}
Dimpal Janu received her B.Tech. degree in Electronics and Communication Engineering from Govt. Mahila engineering College, Ajmer, India in 2014, and M.Tech. degree in Electronics and Communication Engineering from Malaviya national institute of technology (MNIT), Jaipur, India in 2018.
She is currently pursuing a Ph.D. degree from MNIT Jaipur. Her research
interests include wireless communication, cognitive radio network,
machine learning, and deep learning. She is a member of the IEEE student
branch. 
\end{IEEEbiography}

\begin{IEEEbiography}[{\includegraphics[width=1in,height=1.25in,clip,keepaspectratio]{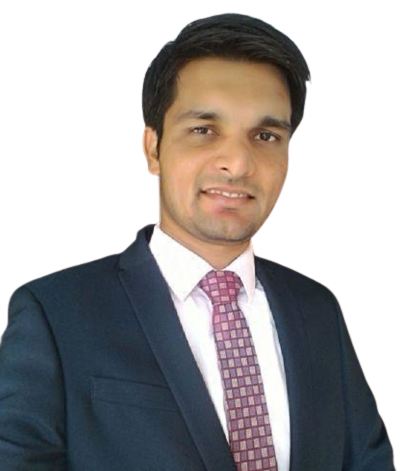}}]{Sandeep Mandia}
received the MTech degree from National Institute of Technology, Silchar, India in 2014 and the PhD degree in computer vision from Malaviya National Institute of Technology Jaipur, India in 2023. He is currently serving as an Assistant Professor at Thapar Institute of Engineering and Technology, Patiala, India. His research interests are machine/ deep learning applications in student engagement analysis, medical diagnostics, and beyond.
\end{IEEEbiography}

\begin{IEEEbiography}[{\includegraphics[width=1in,height=1.25in,clip,keepaspectratio]{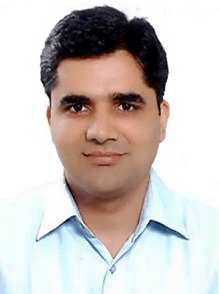}}]{Kuldeep Singh}
received his MTech degree in Signal Processing from Delhi University in $2006$ and a Ph.D. degree in Computer Vision from Delhi Technological University, India, in $2016$. He is currently an Associate Professor with the Department of Electronics $ \& $ Communication Engineering, Malaviya National Institute of Technology, Jaipur, India. Previously, he was a Senior Scientist with the Central Research Lab, Bharat Electronics Ltd., India. He also worked as a postdoctoral fellow at the University of Alberta, Canada from October 2017 to April 2018. His research interest includes Computer Vision, Machine/ Deep Learning, Biometrics, and Cyber Security. He is a reviewer of various IEEE transactions, Elsevier, and Springer journals.
\end{IEEEbiography}
\begin{IEEEbiography}[{\includegraphics[width=1in,height=1.25in,clip,keepaspectratio]{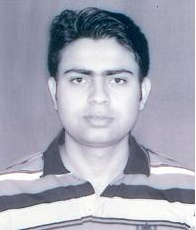}}]{Sandeep Kumar} received his B. Tech. in electronics and
communication from Kurukshetra University, India in 2004 and
Master of Engineering in Electronics and Communication from
Thapar University, Patiala, India in 2007. He received his Ph.D.
from Delhi Technological University, Delhi, India in 2018. He is
currently working as Member (Senior Research Staff) at Central
Research Laboratory, Bharat Electronics Limited Ghaziabad,
India. He has received various awards and certificates of
appreciation for his research activities. His research interests
include the study of wireless channels, performance modeling
of fading channels, and cognitive radio networks. He is also
serving as a reviewer for IEEE, Elsevier, and Springer journals.
\end{IEEEbiography} 
\end{document}